\documentclass[aps,amsmath,amssymb,twocolumn,floatfix,10pt,showpacs]{revtex4}

\usepackage{graphicx}
\usepackage{color}

\begin{document}

\title{ The eight-fold way for optical quasicrystals}
\author{Anuradha Jagannathan}

\affiliation{Laboratoire de Physique des Solides, Universit\'e Paris-Sud, 91405 Orsay, France}
\author{Michel Duneau}

\affiliation{(formerly) Centre de Physique Th\'eorique, Ecole Polytechnique, 91128 Palaiseau, France }

\date{\today}
\pacs{ 67.85.-d, 61.44.Br, 71.10.Fd}

\begin{abstract}
In a recent paper we proposed a means to realize a quasicrystal with eight-fold symmetry by trapping particles in an optical potential created by four lasers. The quasicrystals obtained in this way, which are closely related to the well-known octagonal tiling, offer unique possibilities to study the effects of quasiperiodicity on physical properties.  This method allows, furthermore, to transform the structures, to inflate or deflate them, include interactions or disorder and thus realize a large variety of theoretical models, both classical and quantum. In this paper we present the model, derive a number of interesting geometrical properties of the optical quasicrystals, as well as some results obtained by numerical calculations.
\end{abstract}

\maketitle

\section{Introduction}
The unusual structural properties of solid state quasicrystals are by now well established, more than a quarter century after the first publication \cite{schecht}. The best quasicrystalline samples, which are obtained by standard methods of crystallization, are perfectly ordered over very great distances, as indicated by their extremely narrow, indexable diffraction peaks. In the absence of translational invariance, quasicrystals can have rotational symmetries forbidden for crystals, such as 5-fold, or in the present case, 8-fold symmetry. The property of repetitivity of local environments in quasicrystals ensures that identical finite configurations of atoms occur in close proximity to each other. Quasicrystals have a self-similarity property with respect to change of length-scale. These properties lead one to expect novel physical properties in these materials, and indeed, they are seen to possess intriguing electronic, magnetic and mechanical properties. Unfortunately theoretical understanding of these materials lags behind the experimental findings, in part due to the composition of solid state quasicrystals, which are usually binary or ternary alloys. Their structural complexity makes it impossible to use analytical methods and it stretches numerical computations to the limit. 

Thus, the realization of a single component quasicrystals has been a long-standing goal. 
We recently showed \cite{anumichel} how one can use the optical potential due to four standing wave laser fields to trap particles and realize a two dimensional quasiperiodic structure with eight-fold symmetry. We focus on a particular type of structure, the 8-fold optical quasicrystal, which is closely related to the well-known octagonal  (or Ammann-Beenker) tiling quasicrystal \cite{beenker}.  When the trapped particles are atoms, described by a simple tightbinding model at sufficiently low temperatures, this type of system should provide valuable insights into the quantum properties of quasicrystals. In a quite different context, trapping colloidal particles in such a potential would allow to study phases resulting from competition between the quasiperiodic optical potential and electrostatic interactions.

This paper is organized as follows: Sec.II presents the experimental set-up. Sec.III presents the mathematical background and notation for this problem when lifted to four dimensions. Sec.IV derives the condition that must hold in order to obtain the optical quasicrystal, Sec.V details some of its principal structural properties and Sec.VI discusses some properties of the intensity landscape seen by the trapped particles. Sec.VII presents discussions and conclusions.

\section{Set-up for a quasiperiodic optical potential in two dimensions \label{sec_one}}

\begin{figure}[!ht]
\centering
\includegraphics[width=120pt]{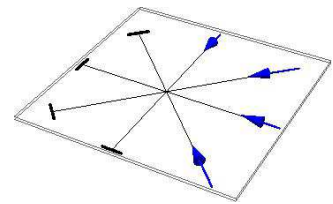}
\caption{Set-up for four standing waves at $\pi/4$ angles using lasers and mirrors. }
\label{laser.fig}
\end{figure}

\begin{figure}[!ht]
\centering
\includegraphics[width=200pt]{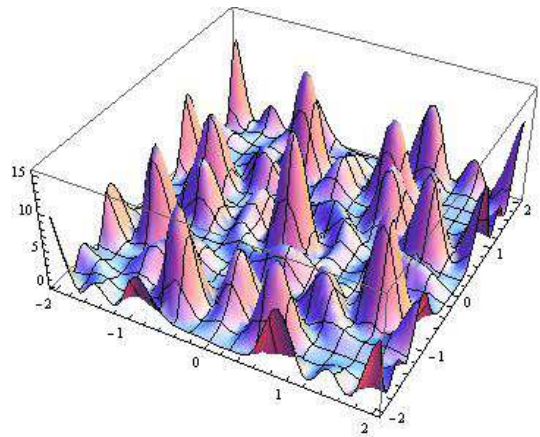}
\caption{Variation of the intensity $I/I_0$ in a region of the $xy$ plane. Distances are expressed in units of $\lambda$.}
\label{pot.fig}
\end{figure}

We consider a region where standing waves have been set up using four laser beams oriented at 45$^\circ$ angles in the $xy$ plane, as shown in Fig.\ref{laser.fig}. The wavelengths, $\lambda$, are the same for all the beams, as are their amplitudes. We will consider the situation where all the polarizations are perpendicular to this plane, allowing the amplitudes to sum up to an absolute maximum. The alternative choice, of using in-plane polarizations yields smaller maxima of amplitudes, smaller contrast, and would therefore be less efficient in trapping particles.
For the case of four standing waves,  the intensity is given by
\begin{equation}
I(\vec{r}) =  I_0 \left[\sum_{n=1}^4 \cos(\vec{k}_n.\vec{r}+\phi_n)\right]^2 
\label{Ixy.eq}
\end{equation}

where $\vec{r}=(x,y)$ is the position vector in the plane. The four wave vectors are given by
\begin{equation}
\vec{k}_n=k(\cos\theta_n,\sin\theta_n) \qquad \qquad \theta_n=\frac{(n-1)\pi}{4}
\label{kvecs.eq}
\end{equation}
with $n=1,..,4$. Notice that the four beams can have arbitrary different phase-shifts $\phi_n$. As long as the relative phase shifts are maintained at some fixed arbitrary values, these phase shifts do not change the nature of the structures obtained, as discussed later.  In colloidal systems, particles are attracted to maxima of the intensity -- the ``optical tweezer`` phenomenon \cite{ashkin}. In cold atomic gases, due to the ac Stark shift, atoms are attracted to the maxima if the laser frequency is chosen to be slightly smaller than the atomic resonance frequency -- so-called ``red'' detuning \cite{bloch}. The optical potential seen by particles in both these situations is $V(\vec{r})\propto -I(\vec{r})$, i.e. particles minimize energy by going towards local maxima of the intensity field. The potential corresponding to Eq.\ref{Ixy.eq} is an instance of a quasiperiodic function, a particular case of almost periodic functions, the theory of which goes back to H. Bohr \cite{bohr} and A. Besicovitch \cite{besic}. Notice that the connection between almost periodic functions and aperiodic sets of points is a far more complex mathematical problem, first 
raised by the mathematician Y. Meyer in the 70's who "invented" the first quasicrystals \cite{meyer}. \\
The plot of Fig.\ref{pot.fig} shows a complex intensity landscape of  local maxima, local minima and saddlepoints with intensities $0 \leq I \leq 16\vert I_0\vert$ obtained for a random choice of the phases $\phi_n$. The function $I(\vec{r})$ is quasiperiodic since there is no integer relationship between the four wave vectors $\vec{k}_n$. The intensities of the peaks, or local maxima,  have a range of values with the upper bound $I_m=16I_0$.

\begin{figure}[!ht]
\centering
\includegraphics[width=150pt]{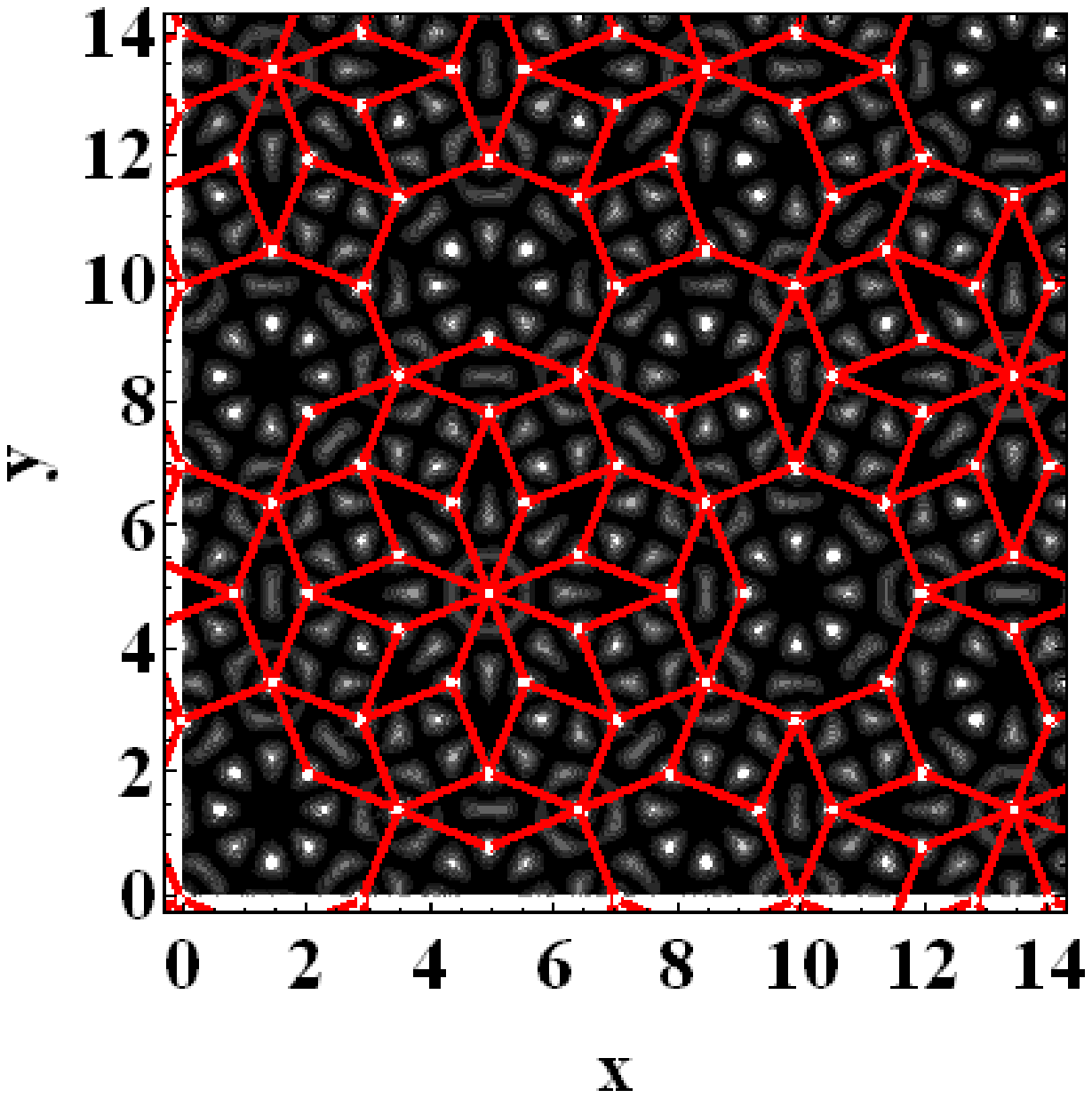}
\includegraphics[width=150pt]{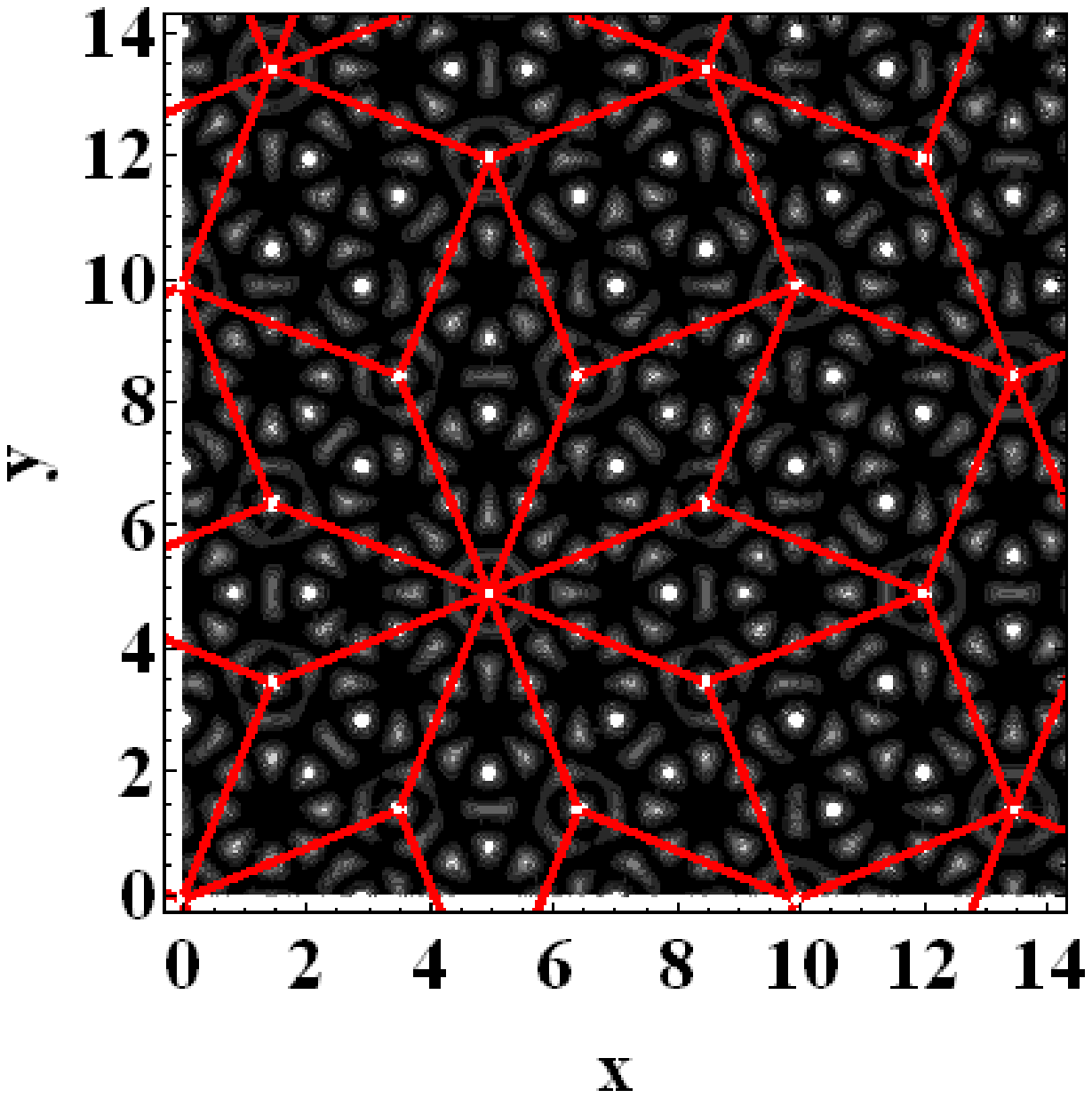}
\caption{Tilings obtained for $I_c/I_0=15$ (top) and $15.82$ (bottom) along with the intensities (represented by a shaded contour plot, darker shade for low intensity).  }
\label{twotilings.fig}
\end{figure}

We now assume that occupied sites are those local maxima for which $\vert I(\vec{r}_j)\vert \geq I_c$.  $I_c$ is the cutoff value, which can depend for instance on the temperature, for systems dominated by thermal fluctuations. In practice, the cutoff is not expected to be sharp, but the corresponding structures are not significantly affected, providing it lies within a sufficiently narrow range of values, as we will see later. This occupancy assumption supposes that there is ergodicity and that particles can effectively reach the minima of the potential. For certain choices of the cutoff, one obtains the kind of tiling  illustrated in Fig.\ref{twotilings.fig}. In Sec.\ref{construction_sec}, we derive this series of values of cutoff, $I_c/I_0$  in terms of wavelength $\lambda$. \\

Figs.\ref{twotilings.fig} show the optical quasicrystals obtained for  $I_c/I_0=15$ (top) and $15.82$ (bottom). The edgelength of the two quasicrystals are approximately $3.81 \lambda$ and $ 9.19 \lambda$ respectively. The tilings are shown superposed on the intensity profile, represented by a shaded plot (dark shades for small intensity). The two are examples of the type of structure that we will refer to as the optical quasicrystal (OQ). It is closely related to the standard octagonal tiling (OT) \cite{octagonal1,octagonal2} composed of squares and 45$^\circ$ rhombuses. For larger and larger values of the cutoff, approaching $16\vert I_0\vert$, only the largest maxima (corresponding to the lowest energy states) will be occupied, and the  atomic density in the $xy$ plane correspondingly decreases. As the cutoff takes on successive values in the series, the lattice vectors get  larger, by powers of the irrational number $\alpha=1+\sqrt{2}$, called the silver mean.

\section{Definitions of the 4D model \label{fourD_sec}}
The wave vectors of the laser beams, $\vec{k}_n$, can be regarded as projections in the $xy$ plane of four orthogonal 4-dimensional vectors $\vec{K}_n=K\vec{\varepsilon}_n$, where $\vec{\varepsilon}_n$ form the standard orthonormal basis. We have $\vec{k}_n = \pi (\vec{K}_n)$, with $K=\sqrt{2}\, k$ and
\begin{equation}
\boldsymbol{\pi}=\frac{1}{2}\left[\begin{smallmatrix}
1 & \frac{1}{\sqrt{2}} & 0 &\frac{-1}{\sqrt{2}} \\
\frac{1}{\sqrt{2}}   & 1  & \frac{1}{\sqrt{2}}  & 0\\
0 & \frac{1}{\sqrt{2}}  &  1  &\frac{1}{\sqrt{2}} \\
\frac{-1}{\sqrt{2}}  & 0 & \frac{1}{\sqrt{2}}  & 1 
\end{smallmatrix}\right]
\end{equation}
is the orthogonal projection matrix onto the $xy$ plane which is denoted by $P$ (the physical plane). The directions orthogonal to $P$ define a perpendicular plane $P'$, so that the 4D space is the direct sum $P\oplus P'$ of the two planes. The corresponding complementary projection matrix is 
\begin{equation}
\boldsymbol{\pi}'=\frac{1}{2}\left[\begin{smallmatrix}
1 & \frac{-1}{\sqrt{2}} & 0 &\frac{1}{\sqrt{2}} \\
\frac{-1}{\sqrt{2}}   & 1  & \frac{-1}{\sqrt{2}}  & 0\\
0 & \frac{-1}{\sqrt{2}}  &  1  & \frac{-1}{\sqrt{2}} \\
\frac{1}{\sqrt{2}}  & 0 & \frac{-1}{\sqrt{2}}  & 1 
\end{smallmatrix}\right].
\end{equation}

\begin{figure}[!ht]
\centering
\includegraphics[width=240pt]{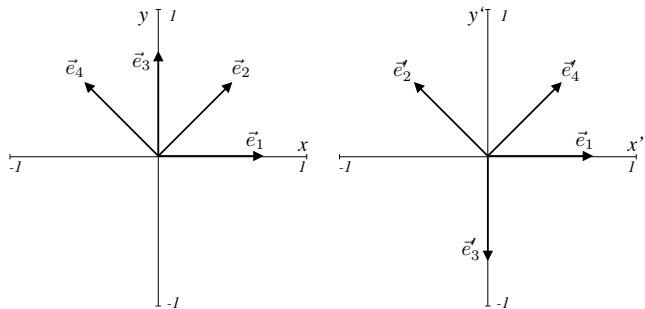}
\caption{Projections  $\vec{e}_n$ in $P$ (left), and  $\vec{e}'_n$ in $P'$ (right).}
\label{evecs_fig}
\end{figure}

We set $\vec{e}_n=\pi (\vec{\varepsilon}_n)$ ($n=1,..,4$)  for the projections of the four 4D basis vectors in the plane $P$. Similarly, we set $\vec{e}_n'=\pi'(\vec{\varepsilon}_n)$  for the projections of the basis vectors in the plane $P'$.   The norms of these vectors, all equal, are $ e_n = e_n'= \frac{1}{ \sqrt{2}}$. In $P$, we choose the orientations of the $x$ and $y$ axes parallel to the vectors $\vec{e}_1$ and $\vec{e}_3$. In the $P'$ plane, the $x'$ and $y'$ axes are oriented along the $\vec{e}'_1$ and $-\vec{e}'_3$ directions, respectively.  These projections are shown in Figs.\ref{evecs_fig}. 

A point $\vec{R}=(R_1,R_2,R_3,R_4)$ in the four dimensional space can be written in terms of its components in $P$ and $P'$ as $\vec{R}=(\vec{r},\vec{r}')$, where $\vec{r}=(x,y)$ and  $\vec{r}'=(x',y')$ lie in $P$ and $P'$ respectively. The transformation can be written in terms of a 4D rotation $\mathcal{R}$ as follows:

\begin{align*}
\left[\begin{array}{l}
x\\
y\\
x'\\
y'
\end{array}\right] &=\mathcal{R}
\left[\begin{array}{l}
R_1\\
R_2\\
R_3\\
R_4
\end{array}\right]
=\frac 1 2
\left[\begin{array}{rrrr}
\sqrt{2} & 1 & 0 & -1\\
0 & 1 & \sqrt{2} & 1 \\
\sqrt{2} & -1 & 0 & 1 \\
0 & 1 & -\sqrt{2} & 1
\end{array}\right].
\left[\begin{array}{l}
R_1\\
R_2\\
R_3\\
R_4
\end{array}\right], \\
\left[\begin{array}{l}
R_1\\
R_2\\
R_3\\
R_4
\end{array}\right]&=\mathcal{R}^t
\left[\begin{array}{l}
x\\
y\\
x'\\
y'
\end{array}\right]
=\frac 1 2
\left[\begin{array}{rrrr}
\sqrt{2} & 0 & \sqrt{2} & 0\\
1 & 1 & -1 & 1 \\
0 & \sqrt{2} & 0 & -\sqrt{2} \\
-1 & 1 & 1 & 1 
\end{array}\right]
\left[\begin{array}{l}
x\\
y\\
x'\\
y'
\end{array}\right] .
\end{align*}
If $\vec{r}'=0$ the last equation gives the 4D coordinates $(R_1,...,R_4)$ of a point $\vec{r}=(x,y)$ in $P$, a property which will be used later on in the analysis of the optical potential. Some of the properties of the 4D cubic lattice with regard to symmetry under eight-fold rotations are given in Appendix A.

\begin{figure}[!ht]
\centering
\includegraphics[width=200pt]{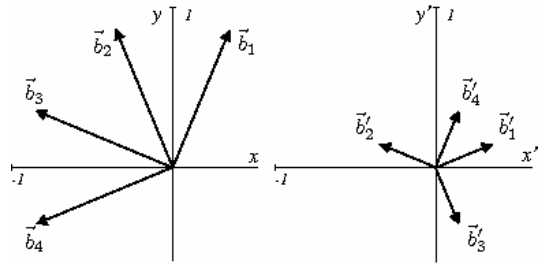}
\caption{Projections  $b_n$ in $P$ (left), and  $b'_n$ in $P'$ (right).}
\label{bvecs_fig}
\end{figure}

We next define the body centered lattice $B$ in 4D, obtained by including the body centers of the 4D hypercubic lattice. 
The projections of the basis vectors of $B$ in the physical and perpendicular planes are $\vec{b}_n=\boldsymbol{\pi}(\vec{\beta}_n)$ and $\vec{b}'_n=\boldsymbol{\pi}'(\vec{\beta}_n)$ for the projections in $P$ and $P'$ respectively. Using the definitions given in Appendix A, one can check that
the angles $(\vec{b}_n,\vec{b}_{n+1})$ are $\frac \pi 4$ and the angles $(\vec{b}'_n,\vec{b}'_{n+1})$ are $\frac {3\pi} 4$  for $n=1,2,3$. One can also check that the norms are
$\beta_n=1$, $b_n=\frac  1 2 \sqrt{2+\sqrt{2}}$ and $b'_n=\frac  1 2 \sqrt{2-\sqrt{2}}$ so that 
$\frac{b_n}{b'_n}=1+\sqrt{2}$. The projected vectors are shown in Figs.\ref{bvecs_fig}. Like the simple cubic lattice, $B$  is invariant under the octagonal group $C_{8v}$. With these definitions, we turn to the description of the experimental set-up and the properties of the optical quasicrystal in the next section.

\vskip 2cm

\section{Construction of an optical quasicrystal \label{construction_sec}}
If $\vec{R}=(\vec{r},\vec{r}')$ and $\vec{K}=(\vec{k},\vec{k}')$ denote the decompositions of 4D vectors $\vec{R}$ and $\vec{K}$ in the direct sum $P\oplus P'$, the 4D scalar product writes $\vec{K}.\vec{R}=\sum K_n R_n=\vec{k}.\vec{r}+\vec{k}'.\vec{r}'$ by orthogonality of $P$ and $P'$.

Let us consider the function $\mathcal{I}$ defined in the 4D space $\mathbb{R}^4$ by
\begin{eqnarray}
\mathcal{I}(\vec{R})&=&I_0\left[ \sum_{n=1}^4 \cos(\vec{K}_n.\vec{R})\right]^2 \nonumber \\
&=&I_0\left[ \sum_{n=1}^4 \cos(\vec{k}_n.\vec{r}+\vec{k}'_n.\vec{r}')\right]^2,
\label{I4d.eq}
\end{eqnarray}
where $\vec{K}_n=\sqrt{2}k \, \vec{\varepsilon}_n=(\vec{k}_n,\vec{k}'_n)$.  

The central point of this discussion is that $I$, the intensity function of Eq.\ref{Ixy.eq}, can be seen to be the restriction to the plane $P$ of the 4D-periodic function $\mathcal{I}$:
\begin{align*}
I(\vec{r})&=\mathcal{I}(\vec{r},0).
\end{align*}

The function $\mathcal{I}$ is periodic with respect to the lattice $\frac{\pi\sqrt{2}}{k} B$ where $B$ is the body centered cubic lattice defined in the previous section. $\mathcal{I}$ has maxima $\mathcal{I}(\vec{R})=16I_0$ on the vertices of the rescaled BCC lattice $\frac{\pi\sqrt{2}}{k} B$. The minima $\mathcal{I}(\vec{R})=0$ correspond to the 3-dimensional hypersurface of equation $ \sum_{n=1}^4 \cos(\vec{K}_n.\vec{R})=0$. This is periodic with respect to BCC lattice, and it intersects $P$ along lines where 
$I(\vec{r})=0$. \\
In the equations above we have not written the phases, for simplicity, as they do not modify the type of structures -- more precisely, adding the phases $\phi_n$ in Eq.\ref{Ixy.eq} is equivalent to a global 4D translation $\vec{\tau}$, such that $K \vec{\tau}.\vec{\varepsilon}_n=\phi_n$. Such translations yield OQ's belonging to the same "local isomorphism class"  (meaning that any finite pattern occurs in both tilings with the same probability). 

\begin{figure}[!ht]
\centering
\includegraphics[width=200pt]{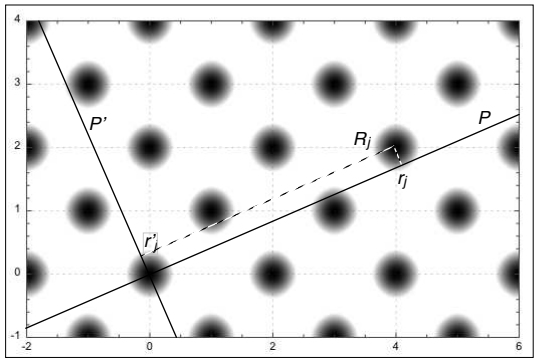}
\caption{The domains where $\left(\cos\pi x+\cos\pi y\right)^2>1.8$ are located on a body centered square lattice. }
\label{fig_three}
\end{figure}

Local maxima of $I$ in the $xy$ plane are ``selected" if their intensity $I(\vec{r})>I_c$ for some value of the cutoff $I_c$. 
For sufficiently large values of the cutoff, the regions in which $\mathcal{I}(\vec{R})>I_c$ form small disjoint domains, $\Delta_{j}$, centered on the lattice points of the BCC lattice, as shown in the 2-dimensional example of Fig.\ref{fig_three}. Therefore a selected local maximum of $I$ occurs when a domain 
$\Delta_j$ intersects the $xy$ plane $P$. 
Expanding around each maximum to quadratic approximation, one finds 
\begin{equation}
\mathcal{I}(\vec{R})/I_0\approx 16-8k^2\vec{R}^2=16-8k^2(\vec{r}^2+\vec{r}'^2)
\label{quad.eq}
\end{equation}
As $I_c$ tends to 16, these domains shrink to lattice points and their shapes tend to spheres of radius $\rho$ such that $8k^2\rho^2=16-I_c/I_0$. In the same limit the projections of $\Delta_{j}$ on $P$ (or $P'$) tend to disks $D_{j}$ (or $D'_{j}$) of radius $\rho$.  \\
If such a domain $\Delta_j$ cuts the plane $P$, the local maximum of $I$ occurs at a point $\vec{s}_j=(x,y)$ which converges to the projection $\vec{r}_j$ of $\vec{R}_j$ in the limit $I_c\rightarrow16$. It is easy to see that
 if $I_c$ is large enough,  the local maxima of $I$ in the domain $I>I_c$ can be indexed by lattice points of the BCC lattice. This point is explained in more detail in Appendix A.

\begin{figure}[!ht]
\centering
\includegraphics[width=150pt]{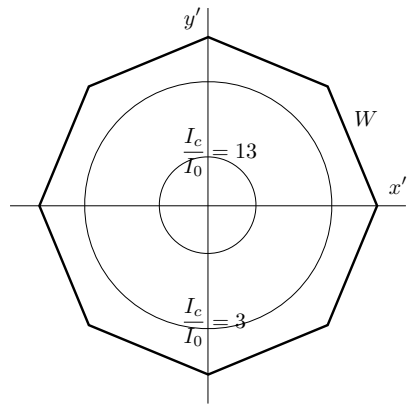}
\caption{Contours of constant intensity in the $P'$ plane for $I_c=3 I_0$ and $13I_0$. The octagonal domain $W$ corresponds to the projection of the BCC unit cell. }
\label{disk.fig}
\end{figure}

One now has an alternative form of the condition which must be satisfied for a lattice point to be selected.  In order for a domain $\Delta_j$, centered about $\vec{R}_j$, to intersect the plane $P$ the projection $\vec{r}'_j$ of $\vec{R}_j$ must lie within a corresponding domain in the $P'$ plane.
 Fig.\ref{disk.fig} shows two domains, for values $I_c/I_0 = 3$ and 13, lying inside an octagon shaped domain $W$, the projection in $P'$ of the BCC unit cell based on the four vectors $\frac{\pi\sqrt{2}} k \vec{\beta}_n$.  For large enough $I_c$, this domain tends to a disk (see Fig.\ref{disk.fig}),  of radius $\rho=\frac {\sqrt{2}} {4k} \sqrt{16-I_c/I_0}$. 

\begin{figure}[!ht]
\centering
\includegraphics[width=150pt]{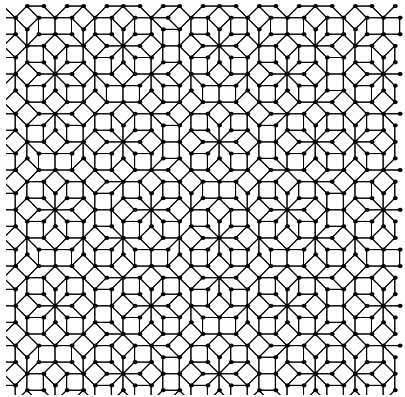}
\caption{A portion of the standard octagonal tiling as obtained by the projection method using the window W.}
\label{standardOT.fig}
\end{figure}

We now turn to the standard octagonal tiling  \cite{octagonal1,octagonal2,octagonal3,ogu,octagonal4}. This tiling, illustrated in Fig.\ref{standardOT.fig}, can be obtained from the four dimensional lattice $\frac{\pi\sqrt{2}} k B$ by the cut and project method. In this case, the selection window is the already defined domain $W$. This tiling can be transformed into itself under a scale change or inflation. This corresponds to defining new tiles which are bigger than the old ones by a factor $\alpha$. This operation can be repeated infinitely many times, and it can be reversed (deflations). Inflations can be carried out in the cut and project method by redefining the selection window in the $P'$ plane. Thus, the series of inflated octagonal tilings are associated with concentric selection windows, $W_1, W_2,...W_p$, the series of octagons scaled down by factors of $\alpha^p$ with $p=1,2,...$. The condition for maximal coincidence of the optical quasicrystal and the octagonal tilings is that the disk $D'$ overlaps with one of these selection windows, $W_p$. The condition for equal areas reads

\begin{equation}
\left|D_p'\right| = \left|W_p\right|  \\
=\frac{\pi^2 \sqrt{2}}{k^2} \alpha^{-2p} ,
\label{condition.eq}
\end{equation}
where we substituted the expression for the area of the octagonal window $|W|= 2\alpha\frac{2\pi^2}{k^2} |b'_n|^2$. This relation gives the size of the domains corresponding to the optical quasicrystal. 

\begin{figure}[!ht]
\centering
\includegraphics[width=240pt]{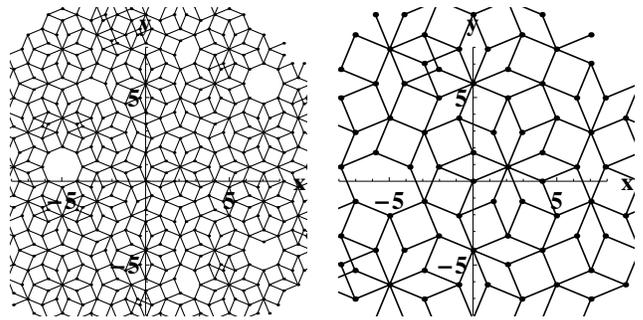}
\caption{Tilings obtained by the projection method using the windows $D'_0$(left) and $D'_1$ (right). $x$ and $y$ are given in units of $\lambda$. }
\label{projected.fig}
\end{figure}

In our present case, with the standard projection method, which consists of substituting $p=0$ in Eq.\ref{condition.eq}, and projecting points whose coordinates $\vec{r}'$ fall within the domain $D_0'$, one obtains the tiling shown in Fig.\ref{projected.fig}(left). The next tiling corresponds to $p=1$ and it is shown in  Fig.\ref{projected.fig}(right). 

The edges of the tiling in $P$ have lengths given by

\begin{equation}
\ell=\frac{\pi \sqrt{2}}{k}\,|b_n| \\
=\frac{\pi}{k} \sqrt{1+\frac 1 {\sqrt{2}}}
\end{equation}

Eq.\ref{condition.eq} can now be used to determine the  cutoff value of the intensity $I_c$. 
This results in the following condition for the special values of $I_c$ , namely

\begin{equation}
I_c/I_0 \approx 16 - 8\sqrt{2}\pi\alpha^{-2p}.
\label{condition2.eq}
\end{equation}

where the quadratic approximation used is satisfactory for $p>1$. The values of $I_c$ for the first few values of $p$ along with the corresponding edge lengths are given in the table below. Note that the tiling corresponding to $p=0$ cannot be obtained by selecting peaks based on their intensities.

\bigskip
\begin{center}
\begin{tabular}{|c|c|c|}
\hline 
 $\makebox[1.cm]{p}$ & \makebox[1.5cm]{$ I_c/I_0$} &  \makebox[1.5cm]{$\ell(\lambda)$} \\
\hline 
0 & N.A. & 0.65 \\
1 & 10.8 & 1.58 \\
2 & 15.0 & 3.81 \\
3 & 15.82 & 9.19 \\
\hline 
\end{tabular}
\end{center}
\vskip 0.5cm {\small{Table I. Values of $p$ and corresponding intensity cutoff values, along with the edge lengths of the optical quasicrystal.}} \\

Using the selection rule $I\geq I_c$ in real space, as opposed to the projection method already illustrated, one gets the structures shown in Figs.\ref{twotilings.fig}. These tilings differ from the ones shown in Fig.\ref{projected.fig}  in one respect which is not visible to the eye, namely the positions of vertices are very slightly displaced from their ideal positions, since they are given by the intersection of physical plane $P$ with the domains $D'$ of the 4D lattice. The displacements from the ideal structure get progressively smaller as the cutoff (or equivalently, $p$) is increased. Another point to note is that all the OQ for $p\geq 1$, can be generated by the intensity selection rule, but not the one corresponding to $p=0$, which can $only$ be obtained by the projection method. This is due to the fact that $\mathcal{I}(\vec{r})$ is a monotonically decreasing function of $\vec{r}$ only up to a certain distance, making the 4D construction inapplicable for small $I_c$. As $I_c$ decreases, the disjoint domains $\Delta_j$ centered on the BCC lattice expand and finally become connected.

To resume, the optical quasicrystals can be obtained in two ways. The usual cut-and-project method using circular windows as the selection rule gives quasiperiodic structures with well defined, fixed, edge lengths. The alternative method is the one we have described using the optical potential and the criterion $I\geq I_c$, which, via Eq.\ref{condition.eq} generates an almost perfect optical quasicrystal. The second method can be experimentally realized in a conceptually simple and physically transparent way, in contrast to the former, which relies on an unphysical 4D theoretical construction.

We conclude this section by noting that the OQ is stable under fluctuations of value of the cutoff in the following sense:  as long as the changes are small compared to the spacing between cutoffs $I_c$, the vertices OQ can be labeled by the set of quasilattice vectors $\vec{r}_n$  for a given value of $p$. Additional vertices appear, in the empty plaquettes for example, if the cutoff is lowered below the ideal value (vertices disappear if the cutoff is raised) without modifying the ``backbone" of the structure.

\section{Structural properties  \label{structure_sec}}
\begin{figure}[!ht]
\centering
\includegraphics[width=120pt]{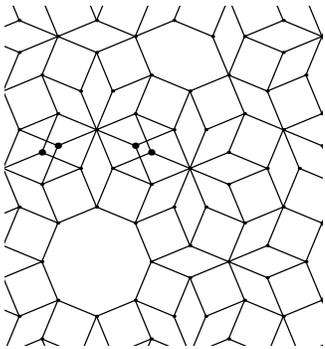} 
\caption{ A close-up of the optical quasicrystal showing the two differences from the OT:  empty hexagon and octagon, and two pairs of "twin" sites (shown by bold dots).}
\label{fig_OQ}
\end{figure}

\begin{figure}[ht]
\centering
\includegraphics[width=120pt]{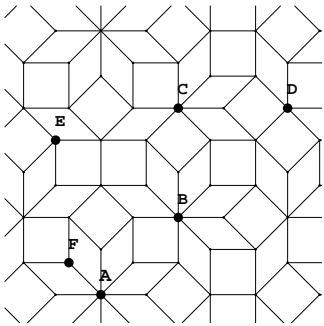} 
\caption{ The octagonal tiling, with sites labeled A,B,... corresponding to coordination numbers $z=8,7,...$.}
\label{fig_OTb}
\end{figure}

In this section we discuss the structure of the optical quasicrystal, compared with that of the standard octagonal tiling. As seen in the previous section, we defined the OQ as the set of points corresponding to a circular selection window of area equal to one of the series of octagons, $W$,$W_1$,.. corresponding to the inflations of the octagonal tiling. The OQ and the OT coincide, for the majority of points, whose perpendicular space projections lie in the intersection of the two selection windows. The differences arise due to the few remaining points. If we take the reference structure to be the OT, then certain points are missing, in the OQ, as can be seen from the empty octagons and hexagons in  Fig.\ref{fig_OQ}, which are not present in Fig.\ref{fig_OTb}. Conversely, there are some closely-spaced sites in the OQ, not present in the OT. A simple calculation shows that about 2\% of the OQ points are 
missing in the OT and the same percentage of the OT points are missing in the OQ. 

\begin{figure}[ht]
\centering
\includegraphics[width=180pt]{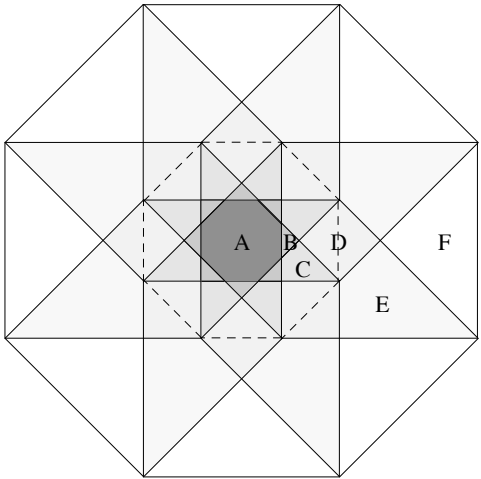} 
\caption{ The octagonal tiling selection window in $P'$ indicating the subwindows for each family, corresponding to the labelling in Fig.\ref{fig_OTb}. The axes have been rotated by $\pi/8$ with respect to Fig.\ref{disk.fig}.}
\label{fig_OTa}
\end{figure}

To understand more precisely these structure differences between the OT and the optical quasicrystal, it is necessary to enter into the details of the perpendicular space projections of different local environments of the octagonal tiling. These local environments correspond to coordination numbers $z=8,7,...,3$ and  are labeled A,B,C,..,F as shown in  Fig.\ref{fig_OTb}. It can be shown \cite{octagonal4} that the coordination number $z$ of any vertex of the OT is simply related to its position in $P'$, as shown in  Fig. \ref{fig_OTa}. Comparing the two acceptance domains for the OT and for the OQ, one sees that the differences arise in the outlying E and F domains. The resulting differences are that  i)  some sites are missing, giving rise to the empty octagons and hexagons in the OQ, and ii)  some new F sites appear in close proximity to another F site. In quasicrystals such as the OT,  two such close-lying sites are well-known in the literature, as being related by a ``phason-flip". A phason excitation mode is a collective excitation made up of many such individual hops \cite{hen}. Whereas in quasicrystals only one of the pair of sites is occupied, in the present case of the OQ,  both sites are simultaneously occupied -- we will refer to these as twin-sites. 
Thus, in sum, the OQ has some missing sites, leading to holes in the structure, and some twin-sites, compared to the OT which is a more homogeneous structure. These structure differences, which concern a small fraction of atomic positions, differentiate the OQ from the standard OT.  In the next section we indicate how these differences can be done away with or minimized, by introducing repulsive interactions.

One can now understand the effects of small changes in the cutoff intensity with respect to the values given by Eq.\ref{condition2.eq}. When the value of the cutoff $I_c$ is slightly higher than the value given by Eq.\ref{condition.eq} corresponding to a particular OQ, some points are eliminated, leading to a structure with more holes. Conversely, if the cutoff is lowered slightly with respect to the optimal value then the density of points slightly increases, along with the appearance of more twin-sites. One can introduce disorder in the model by smoothing the cutoff, which we have assumed to be abrupt. If the selection condition is relaxed, by imposing a  stochastic rule for peaks of intensities $I$ close to $I_c$, there will be a spatial disorder. One can smooth the irregularities in spatial density by introducing repulsive interactions (next section).

\vskip 2cm

We now turn to the diffraction pattern of the optical quasicrystal, one of its readily measurable properties in an experiment.  The general method for finding the Fourier transform and indexation of peaks is discussed for example by A. Katz and D. Gratias in \cite{diffrac}. The structure factor consists of Bragg peaks whose positions are given by the reciprocal lattice of the 4D cubic lattice, and whose peak intensities are given by the Fourier Transform of the selection window. Due to the close similarity of their selection windows, the structure factor of the OQ and that of the OT are expected to be very similar, with peaks in the same positions but with slight differences in their intensities.  Here we will confine our attention to the indexing of Bragg peaks.  It is interesting to observe that spherical selection windows were introduced by Grimm and Baake in the context of the diffraction patterns of quasiperiodic tilings \cite{grimmbaake}, as a useful approximation in the calculation of the structure factor of, for example, the icosahedral tiling. For the optical quasicrystal, in contrast, we see that the spherical window turns out to be the correct, physically imposed choice, provided only that the cut-off $I_c$ is large enough. \\

The 4D basis vectors are $\vec{K}_n = (2\sqrt{2} \pi/\lambda) \vec{\varepsilon}_n$. After projection in $P$, one obtains the four lattice vectors of Eq.\ref{kvecs.eq}
\begin{equation}
\vec{k}_n = \frac{2\pi}{\lambda}\vec{e}_n,
\end{equation}
 The peaks of the structure factor $S(\vec{k})$ are found at positions $ \vec{k}=\sum h_n \vec{k}_n$ with the condition that
\begin{equation}
\exp(i \pi \sum_n h_n) = 1
\label{hcond.eq}
\end{equation}

\begin{figure}[!ht]
\centering
\includegraphics[width=180pt]{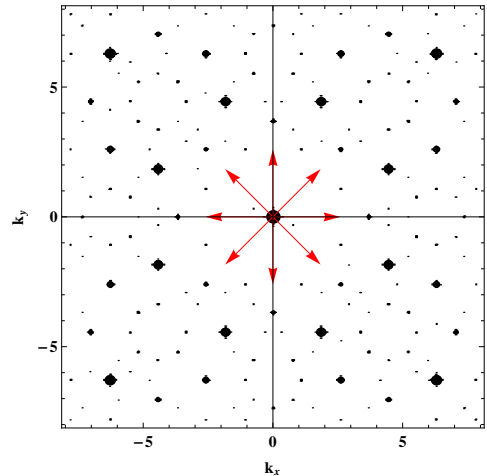}
\caption{ The structure factor calculated using Eq.\ref{struc.eq} for a sample of 4000 sites of the optical quasicrystal (p=0). Intensities of the peaks are proportional to the area of the spots, $k_x$ and $k_y$ are given in units of $\lambda^{-1}$. Arrows indicate the eight shortest reciprocal lattice vectors.}
\label{struc.fig}
\end{figure}

This condition is equivalent to saying that the positions of its Bragg peaks are those of $k\sqrt{2}F$, the reciprocal lattice of $\frac{\pi\sqrt 2}{k}B$, after projection into $P$. The values of the intensities are determined by the Fourier transform of the selection window, which is of octagonal shape for the OT, and circular shape for the OQ. The resulting differences between the two structure factors would show up only under a detailed comparison of peak intensities. Fig.\ref{struc.fig} shows the eight vectors $\pm k \vec{k}_n$, and the numerically calculated intensity function $S(\vec{k})$ defined by

\begin{equation}
S(\vec{k}) = \frac{1}{N} \sum_{i,j} \exp(i \vec{k}.(\vec{r}_i-\vec{r}_j)
\label{struc.eq}
\end{equation}
where $N$ is the number of sites in the sample. The set of intense peaks nearest the origin correspond to the combinations $\{\pm1,\pm1,0,0\}$ and permutations thereof, in accordance with the condition in Eq.\ref{hcond.eq}. The structure factors of successive OT for $p\geq1$ will be the same, only rescaled by powers of $ \alpha^{-p}$, to take into account the inflation in real space.

\section{Distribution of intensities in the optical quasicrystal. \label{pot_sec}} 
\begin{figure}[!ht]
\centering
\includegraphics[width=180pt]{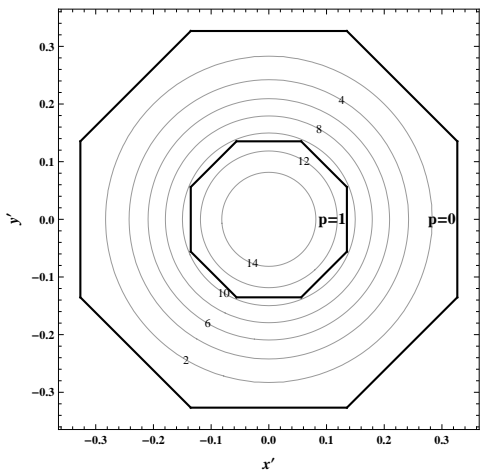} 
\caption{ Contour plot of the intensity function $I_{\perp}(x',y')$ in $P'$. The OT selection windows for $p=1$ and $p=2$ are shown, for comparison. Distances are in units of $\lambda$. The axes have been rotated by $\pi/8$ with respect to Fig.\ref{disk.fig}.}
\label{perppot.fig}
\end{figure}

\subsubsection{Global distribution function of peak intensities}
To describe quasicrystals made by trapping particles in this quasiperiodic potential, one needs to know the distribution of values of their potential energy, $V_i$.  We recall that the potential energy is proportional to the intensity, and that in terms of dimensionless variables $V_i/V_0 = -I(\vec{r}_i)/I_0$. The distribution of potential energies is the distribution of values of the peak intensities, $I(\vec{r}_i)/I_0$, which we denote $P(I)$. The 4D description of the optical potential allows one to calculate $P(I)$ explicitly for large values of the intensity.
Using the 4D function $\mathcal{I}$ defined in Eq.\ref{I4d.eq} we define the intensity function in the $x'y'$ plane, $I_{\perp}(\vec{r}')$:
\begin{equation}
I_{\perp}(\vec{r}')= \mathcal{I}(\vec{r}=0,\vec{r}') = I_0 \left[\sum_{n=1}^4 \cos(\vec{k}'_n.\vec{r}')\right]^2 
\end{equation}
where
\begin{equation}
\vec{k}'_n=k(\cos\theta_n,\sin\theta_n) \qquad \qquad \theta_n=\frac{3(n-1)\pi}{4}
\label{kpvecs.eq}
\end{equation}

\begin{figure}[!ht]
\centering
\includegraphics[width=200pt]{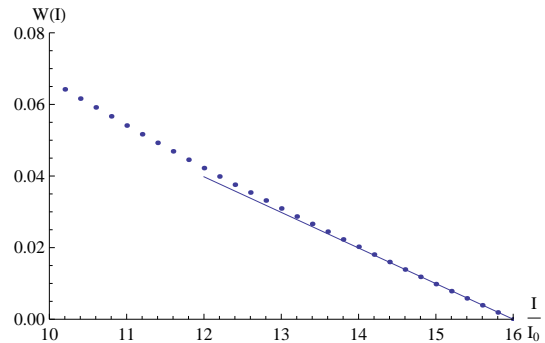} 
\caption{ Integrated distribution function for peak intensities, $W(I)$. Points represent averages over many samples (see text). The line corresponds to the theoretical form in Eq.\ref{probdist.eq}. }
\label{probdist.fig}
\end{figure}

The intensity of a peak, $I(\vec{r}_i)$ is given to good approximation by $I_{\perp}(\vec{r}')$. Fig.\ref{perppot.fig} shows the constant intensity contours of this function in the $P'$ plane, along with the selection windows $W_1$ and $W_2$. 
The total number of peaks of intensity between $I$ and $I_m$,  $N(I) \sim \int_I^{I_0} P(I) dI$, is proportional to the area $A(I)$ enclosed by the contour $I_{\perp}(x',y')=cst$. For large intensity,  the quadratic approximation for $I_{\perp}$ holds, and the contours are circles. For large $I$, one has

\begin{eqnarray}
N(I) &=& A(I) \nonumber \\
&=&  \frac{1}{32\pi} \left(I_m-I\right)
\label{probdist.eq}
\end{eqnarray}
This results in a probability distribution $P(I) \sim C$ for large $I$, where $C$ is a normalization constant. We have compared this prediction with the peak intensity distribution calculated numerically. Samples were created by taking a variety of random angles $\phi_n$ in Eq.\ref{Ixy.eq}, and the largest peaks were found numerically. The sample averaged probability distribution $P(I)$ and its integral $N(I)$ were then calculated.  Fig.\ref{probdist.fig} shows the resulting numerically determined values of $N(I)$ for $10I_0 \leq I \leq 16I_0$, along with the straight line behavior predicted by Eq.\ref{probdist.eq}. One finds that there is good agreement for large intensities, as expected. For smaller intensities, the curve departs from the straight line and rises faster. This is in part due to the deformation of the contours due to the fourth (and higher) order terms which break the rotational invariance.  It is interesting to note that for an optical potential with decagonal symmetry but not of the standing wave type we consider here, the probability distribution of peak heights was found to be a power law, $P(I) \sim (I/I_m)^{-3/4}$ \cite{schmied}.

\subsubsection{Onsite potentials for different families of sites.}
\begin{figure}[!ht]
\centering
\includegraphics[width=240pt]{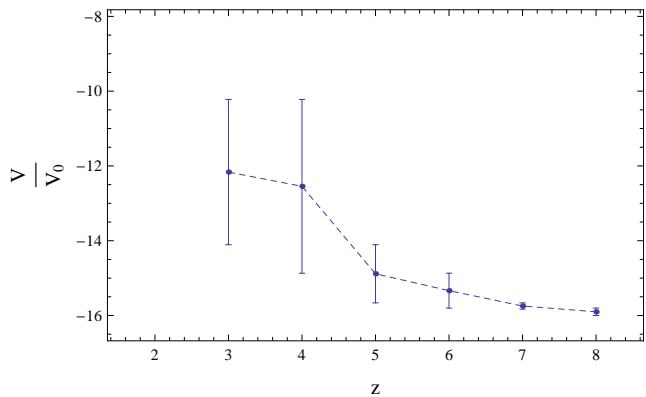} 
\caption{Maximum and minimum values of the onsite potential plotted versus coordination number $z$ for the $p=1$ tiling. The dashed line is a guide to the eye.}
\label{histo.fig}
\end{figure}

We next discuss the onsite potential, as a function of the type of site. As was seen in Sec.\ref{structure_sec}, each of the families of sites A,B,... corresponds to a domain in $P'$. The points near the origin, which correspond to A sites of the OT, have large values of the intensity, and the intensity diminishes as one goes to the B, C, .., out to the F sites. Thus, the peak intensity is decreasing as one goes from $z=8$ (A sites) down to $z=3$ (F sites). In other words, the $z=8$ sites are located in the deepest potential wells. This statement applies to the mean value of $I$ for each class of sites, while the actual value for a given site will lie within a certain interval around the mean value. Fig.\ref{histo.fig} shows the spread of values of the local intensity corresponding to each value of $z$. The maximum and minimum values of the potential were calculated based on the perpendicular space dependence of the intensity shown in Fig.\ref{perppot.fig} using the domains shown in Fig.\ref{fig_OTa}. This interval grows as $z$ is reduced: the distribution is  narrow for A sites, and gets fairly broad for the E and F sites.

Twin-pairs, as defined in Sec.\ref{structure_sec} are pairs of sites which are close together in real space. In $P'$, they are at opposite ends the selection window, such that their distances from the origin $r'$ are very similar. Since the intensity is a smooth function of $r'$, the difference of peak intensity between two twin F sites will therefore be small. As a consequence of this property, a small repulsive interaction between particles should be sufficient to raise the energy of a twin-pair, making it energetically unfavorable. Conversely, it would become favorable to occupy empty hexagons and octagons, since the empty site in their centers corresponds to an intensity which is just below the cutoff. Thus, in conclusion, small repulsive interactions should induce a transformation of the OQ into the OT. This is an interesting possibility, since there is a large body of work devoted to studying models of electronic and vibrational properties of the OT \cite{sire1,ben,ben2,los,moss2,jagsch,zhong2,jag,trambly}.

\subsubsection{Intensity variations between sites.}
We examine in the section the problem of the behavior of $I(\vec{r})$  between two adjacent sites in the optical quasicrystal. This information is relevant, for example, in discussions of the tunnelling probability between adjacent sites. A priori, it is clear that the variations of the intensity should be different for each pair of sites in the quasicrystal. However, one can find a close similarity of behavior for different types of pairs, depending on their relative positions in the 4D lattice. We describe below the typical intensity profiles for two types of nearest neighbors in a qualitative fashion, leaving a more complete study for a future work. The two kinds of pairs are i) {\it edge-linked pairs} where the distance between sites is $\ell$ (the edge length) and ii) {\it small-diagonal pairs} where the distance between the pair is the length of a small diagonal of the 45$^\circ$ rhombus, $\ell\sqrt{2-\sqrt{2}}$.

The simplest case is that of the densest theoretical structure, the $p=0$ optical quasicrystal (not  realizable using the intensity cut-off method).
\noindent
i) Edge-linked pairs. Moving from site $i$ to site $j$ corresponds, in 4D space, to going along a body diagonal. The function $\mathcal{I}$ must have at least one zero along this path, since the amplitude $\sum_{n=1}^4 \cos(\vec{k}'_n.\vec{r}')$ changes from $-4$ to $+4$. 
After projecting into the physical plane $P$,the intensity would go from a local maximum, through zero, and back up to a local maximum.  \\
\noindent
ii) Small-diagonal pairs. Moving from site $i$ to site $j$ corresponds, in 4D space, to going along an edge of the 4D hypercube. The intensity would evolve from a local maximum to another local maximum without passing through zero in between. \\

\begin{figure}[!ht]
\centering
\includegraphics[width=180pt]{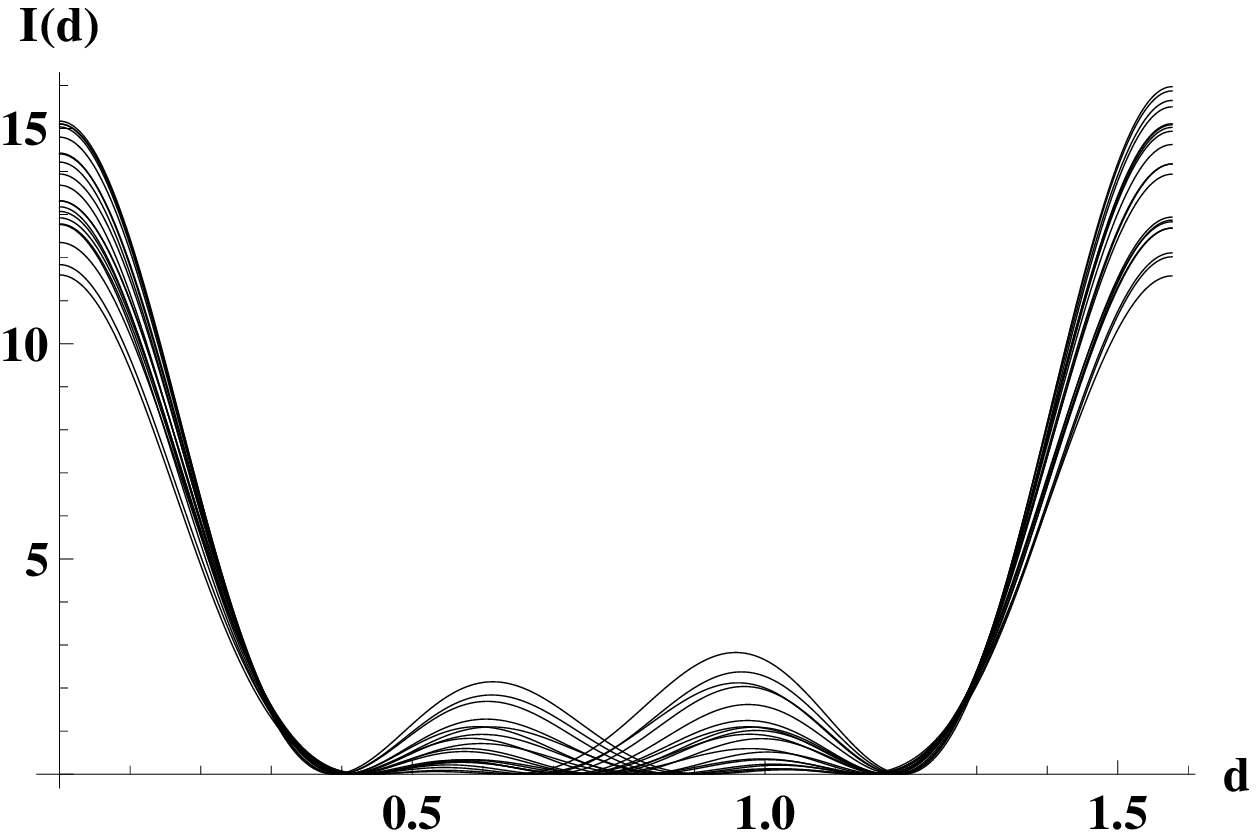} 
\includegraphics[width=180pt]{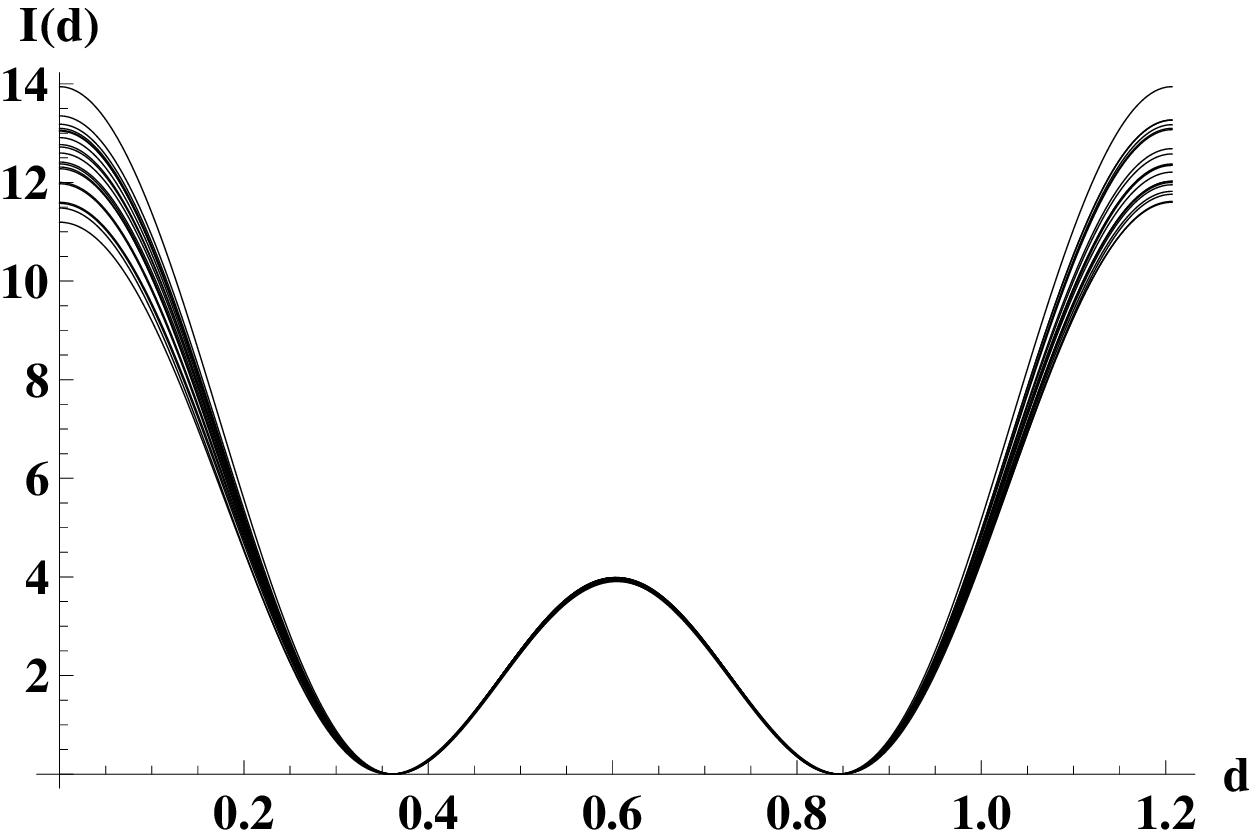} 
\caption{(top) $(p=1)$ Variation of the potential as a function of distance edge-linked pairs (bottom)  $(p=1)$ Variation of the potential as a function of distance for small-diagonal pairs. }
\label{bond1.fig}
\end{figure}

We now consider the physically relevant $p=1$ OQ, where neighboring sites are spaced further apart and belong in different unit cells in 4D. This leads to intensity variations which have a more complicated dependence upon distance. There are additional maxima and minima between the start and finish, compared to the simple case discussed above. However there is still a strong similarity between the curves obtained for each of the two cases i) sites joined by an edge and ii) sites lying across the small diagonal. This is now shown by numerical computation, for the case of the optical quasicrystal obtained by imposing the cutoff value $I_c=11$. For pairs joined by an edge, results are shown in Fig.\ref{bond1.fig}a, where the variation of the intensity $I(d)$  is plotted as a function of the distance along the edge $d$ for a large number of pairs. The curves are individually distinct, however they show similar qualitative behavior. A similar set of computations was carried out for sites separated by a short diagonal, with the results shown in  Fig.\ref{bond1.fig}b. These plots indicate that the intensity profile is determined by the type of pair (joined along an edge or along the small diagonal) to first approximation. One can expect therefore the tunnelling amplitude between sites should be approximately of the same order of magnitude for pairs of a given type. A detailed quantitative study to check this statement about the  intensity profiles for the optical quasicrystal is left for future calculations.

\section{Discussion and conclusions. \label{conclu_sec}} 
We have discussed how to obtain an optical quasicrystal with eight-fold symmetry by trapping particles in a laser potential generated by four lasers. By varying temperature for a fixed laser intensity $I_0$ or by varying the laser intensity at fixed temperature, one can obtain structures which are closely related to the standard octagonal tiling, whose properties have been investigated by many authors. The optical quasicrystal can be made to transform into the octagonal tiling by increasing repulsive interactions between particles. The densest optical quasicrystal has an edge length of the order of the laser wavelength, while the infinite series of inflated tilings can be obtained in principle by decreasing the number of particles and increasing the laser intensity. We  presented a four dimensional model and derived the relation between the quasi lattice vectors and the laser wavelength. The model also allows us to calculate the distribution of perak heights analytically for large intensities. We also give the typical values and the widths of the distribution of peak intensities for different families of sites of the optical quasicrystal and we discussed the form of the intensity variations between pairs of sites. This information is relevant to experimental realizations and for numerical studies of such systems. 

The optical quasicrystal, due to its conceptual simplicity, and the numerous experimental tuning parameters that one can use, is an excellent candidate to study the physical properties of quasicrystals. The method described here to obtain an 8-fold quasicrystal can be readily generalized to describe systems with 5-fold and 10-fold symmetry. Quasicrystals made by trapping colloidal particles (typically of nm size) could be used to study melting and phase transitions along the lines of the study in \cite{schmied2}. On a smaller length scale, cold atom quasicrystals would allow to study the effects of quasiperiodicity on the  quantum diffusion of wave packets, and on a variety of transport properties. They offer a new solution to the long-standing problem of experimentally creating and studying single-component quasicrystals.

\acknowledgments
We thank U. Grimm for bringing Ref.\cite{grimmbaake} to our attention.

\appendix
\section{8-fold symmetry of the 4D cubic lattices}
In this appendix, we review briefly some properties relating to 8-fold symmetry, namely, invariance under rotations by $2\pi/n$ with $n=8$. The octagonal group $C_{8v}$ has 16 elements: In 2 dimensions, the 8-fold generator is a rotation $g$ of $\frac{\pi} 4$ or $g'$ of $\frac{3\pi} 4$, giving rise to inequivalent representations of the group. The remaining elements are vertical "mirror" planes which make two conjugation classes. As is well-known, an 8-fold rotational symmetry is not compatible with the translation invariance of a 2D lattice since the trace $tr(g)=\sqrt{2}$ is not an integer as it should be if the matrix of $g$ could be written in the basis of an invariant lattice. However, the group $C_{8v}$ has crystallographic representations in 4D, as we will now explain in more detail. Generators of this representation are an 8-fold element $\gamma$ which we choose as: 
\begin{equation}
\gamma=\left[\begin{matrix}
0&0&0&-1\\
1&0&0&0\\
0&1&0&0\\
0&0&1&0
\end{matrix}\right]
\end{equation}
and a "mirror" plane $\sigma$, 
\begin{equation}
\sigma=\left[\begin{matrix}
1&0&0&0\\
0&0&0&-1\\
0&0&-1&0\\
0&-1&0&0
\end{matrix}\right].
\end{equation} 
They satisfy the following relationships: $\gamma^8=I$, $\sigma^2=I$ and $\sigma\gamma\sigma=\gamma^{-1}$. 
All elements of the group are of the form $\gamma^k$ or $\sigma\gamma^k$ for $0\leq k <8$, the latter are symmetries of order 2. As can be checked readily, the hypercubic 4D lattice $\mathbb{Z}^4$ is invariant under transformation by these matrices and consequently under the whole group. This representation of $C_{8v}$ is reducible: the 4D space $\mathbb{R}^4$ splits into the planes $P$ and $P'$ defined above which are invariant and carry inequivalent irreducible representations of the group. The restriction of $\gamma$ to $P$ is the rotation $g$ of angle $\frac \pi 4$ whereas the restriction to $P'$ is the rotation $g'$ of angle $\frac {3\pi} 4$.\\

We next turn to the body centered lattice $B$. The vertices of $B$ can be written as $\vec{R}=T\vec{n}=\sum j_n\beta_n$ with $\vec{n}=(n_1,n_2,n_3,n_4)$ (where the $n_i$ are integers) and where $T$ is the matrix 
\begin{align*}
T=\frac 1 2
\left[\begin{array}{rrrr}
1 & -1 & -1 & -1 \\
1 & 1 & -1 & -1 \\
1 & 1 & 1 & -1 \\
1 & 1 & 1 & 1
\end{array}\right].
\end{align*}
A basis of $B$ is given by the vectors $\vec{\beta}_n=T\vec{\varepsilon}_n$  ($n=1,..,4$). 
One can check that $\det(T)=\frac 1 2$ and that $T.C_{8v}=C_{8v}.T$ so that $B$ is invariant with respect to $C_{8v}$  and belongs to the same Bravais class as the hypercubic lattice $\mathbb{Z}^4$. \\

We now discuss the indexation of peak positions using the BCC lattice points. When $I_c$ tends to 16$I_0$, the acceptance domains associated with each of the BCC lattice points shrink and their shapes tend to spheres, while their projections on $P$ (or $P'$) tend to disks $D_{j}$ (or $D'_{j}$) of radius $\rho$.  
If such a domain $\Delta_j$ cuts the plane $P$, the local maximum of $I$ occurs at a point $\vec{s}_j=(x,y)$ which converges to the projection $\vec{r}_j$ of $\vec{R}_j$ in the limit $I_c\rightarrow16$.
The 4D coordinates of $\vec{s}_j$  are given by the coordinate transformation 
$\vec{S}_j=\mathcal{R}^t(x,y,0,0)$. $\vec{S}_j$ belongs to $P$ and is close to the lattice point $\vec{R}_j$, so that $\vec{S}_j\approx \frac{\sqrt{2}\pi}{k} T\vec{j}$. 
Thus, if $I_c$ is large enough,  the local maxima of $I$ in the domain $I>I_c$ can be indexed by lattice points of the BCC lattice.

\section{{Expansion of the 4D intensity $\mathcal{I}$ \label{appendix}}}
\label{invpol}
The quadratic approximation Eq.\ref{quad.eq} of the 4D intensity $\mathcal{I}(\vec{R})$ about a
maximum follows from the Taylor expansion of the $\cos$ functions. Within this approximation, the
$C_\infty$ symmetry of the r.h.s. implies that the location of a local maximum of
$I(\vec{r})$ coincides with the projection $r_j$ of the nearest maximum of the 4D function
at the BCC lattice point $\vec{R}_j=(\vec{r}_j,\vec{r}'_j)$. However higher order Taylor expansions
of $\mathcal{I}(\vec{R})$ break this $C_\infty$ symmetry. One can check that the expansion at
order 4 writes
\begin{align*}
\mathcal{I}(\vec{R})&\approx16-8k^2R^2 +\tfrac 3 2(k^2R^2)^2+k^4 r^2r'^2\\
&+k^4\left[ \tfrac 23\left(p_4(\vec{R})+p'_4(\vec{R})\right)+ r_4(\vec{R}) \right],
\end{align*}
where $\vec{R}=(\vec{r},\vec{r}')$.
The $C_{8v}$-invariant polynomials $p_4$, $p'_4$ and $r_4$ are given by
\begin{align*}
p_4(z,z')=\Re[z^3\overline{z}'],  & \qquad p'_4(z,z')=\Re[z\overline{z}'^3],\\
 r_4(z,z')&=\Re[z^2 z'^2],
\end{align*}
where $\vec{r}=(x,y)$, $\vec{r}'=(x',y')$ and $z=x+iy$, $z'=x'+iy'$. \\
Accordingly, a local maximum of $I(\vec{r})$ does not necessarily coincide with the projection
$r_j$ of the nearest maximum of $\mathcal{I}$ at the BCC lattice point $\vec{R}_j$.
The $C_{8v}$ invariance of the 4th-order terms follows from the action of the rotation $\gamma$ and
the symmetry $\sigma$ in the $(z,z')$ representation of $(\vec{r},\vec{r}')$:
$\gamma(z,z')=(\zeta\,z,\zeta^3\,z')$ where $\zeta=e^{i\pi/4}$ and
$\sigma(z,z')=(\overline{z},\overline{z}')$.


\end{document}